\documentclass[10pt,letterpaper,twocolumn]{article} 

\usepackage{ol2}
\usepackage[draft]{hyperref}
\usepackage{amsmath}

\newcommand{\de}{\partial}

\newcommand{\eq}[2]{\begin{equation} \label{#1} #2 \end{equation}}

\begin{document}

\twocolumn[ 

\title{Jackiw-Rebbi states and trivial states in interfaced binary waveguide arrays with cubic-quintic nonlinearity}

\author{Truong X. Tran $^{*}$ and Xuan N. Nguyen}

\affiliation{Department of Physics, Le Quy Don Technical University, 236 Hoang Quoc Viet str., 10000 Hanoi, Vietnam\\
 $^{*}$Corresponding author: tranxtr@gmail.com}

\date{\today}

\begin{abstract}
We systematically investigate two types of localized states - one is the optical analogue of the quantum relativistic Jackiw-Rebbi states, and the other is the trivial localized state - in interfaced binary waveguide arrays in the presence of cubic-quintic nonlinearity. By using the shooting method we can exactly calculate the profiles of these nonlinear localized states. Like in the case with Kerr nonlinearity, we demonstrate that these nonlinear localized states in the case with cubic-quintic nonlinearity also have a distinguishing feature which is completely different from all other well-known nonlinear localized structures in other media. Namely, the profiles of nonlinear localized states with higher peak amplitudes in interfaced binary waveguide arrays can totally envelope those with lower peak amplitudes. We show that high values of the saturation nonlinearity parameter can help to generate and stabilize these intense localized states during propagation, especially in the case with negative coefficient for the cubic nonlinearity term.
\end{abstract}

]

\section{INTRODUCTION}
\label{Introduction}
Waveguide arrays (WAs) present interesting platforms to investigate many fundamental photonic phenomena in classical physics such as discrete diffraction \cite{nature,jones}, discrete solitons \cite{nature,christodoulides,kivshar,agrawal2}, diffractive resonant radiation \cite{truong2}. These platforms have also been used to mimick fundamental effects in nonrelativistic quantum mechanics such as photonic Bloch oscillations \cite{nature,peschel,pertsch,lenz}, and Zener tunneling \cite{ghulinyan}. However, for studying relativistic quantum mechanics phenomena arising from the Dirac equations one needs to use binary waveguide arrays (BWAs) instead of conventional WAs. Recently, several important relativistic quantum mechanics phenomena such as {\em Zitterbewegung} \cite{zitterbewegung}, Klein paradox \cite{klein}, and Dirac solitons in the nonlinear regime \cite{trandirac1}, have been found in BWAs.

The discrete gap solitons in BWAs in the classical context have been explored much earlier, dating back to 1992 \cite{kivshar2,sukhorukov2,conforti11,morandotti,johansson1,gorbach,johansson2}. However, discrete solitons in BWAs were shown to be optical analogues of Dirac solitons (DSs) in a 1D nonlinear relativistic quantum Dirac equation only in 2014 \cite{trandirac1}. The stability, dynamics of these 1D spatial DSs and interaction between them have been studied in \cite{trandirac2}. The higher-order DSs in BWAs have been investigated in \cite{trandirac4}. The 2D spatial DSs in square binary waveguide lattices have been investigated later in \cite{trandirac3}. The Dirac light bullet which can conserve its profile during propagation in both space and time domains has been shown to exist in BWAs in \cite{trandirac5}. The switching of a DS by using an extremely  weak signal in BWAs with varying propagation mismatch (this model can describe the Dirac equation in curved spacetime) has been demonstrated in \cite{trandirac6}. It is worth mentioning that nonlinear Dirac equations have been analyzed since a long time, e.g., by Heisenberg \cite{Heisenberg57}.

In 2017, it was shown in \cite{tranjr} that one can create the optical analogues of special states, well known in the quantum field theory as {\em Jackiw-Rebbi} (JR) solutions \cite{jackiw}, and trivial localized states at the interface of two BWAs. This system is described by two Dirac equations with Dirac masses of opposite signs. The different scenarios of interaction between JR states and DSs in BWAs have been investigated in \cite{tranjrdsinteraction}. Based on the JR states, the charge fractionalisation phenomenon (which is crucial in the discovery of the fractional quantum Hall effect \cite{laughlin}) has been predicted. One of extraordinary properties of the JR states is the topological nature of its zero-energy solution which is interpreted as a precursor to topological insulators \cite{hasan}. Topological photonics can have a great potential in designing robust optical circuits \cite{rechtsman}. In 2019, the JR states in interfaced BWAs were demonstrated, as expected, to be also extremely robust under influence of strong disturbance of various kinds such as the turning on/off of the nonlinearity, the linear transverse potential, and the oblique incidence \cite{tranjr3}.

The above-mentioned JR states and trivial states in \cite{tranjr} have been investigated in the \emph{linear regime} where their exact analytical solutions have been found. The JR states and trivial states in the regime of Kerr nonlinearity have been systematically investigated later in \cite{tranjr2} both for self-focusing and self-defocusing nonlinearity by using the shooting method \cite{shooting}. This model with Kerr nonlinearity is the simplest one for studying third-order nonlinear effects in optics. However, if the optical signals are intense enough one needs to take into account the fifth and even higher-order terms for nonlinearity. The resulting equation in that case is often called the cubic-quintic nonlinear Schr\"{o}dinger equation (NLS) because it contains terms accounting for both the third and fifth powers of the signal amplitude. In fiber optics this cubic-quintic NLS for a \emph{single} fiber has been well studied \cite{akhmediev}.

In this work, we study the JR states and trivial states in interfaced BWAs with cubic-quintic nonlinearity. The paper is organized as follows. In Sections 2, as a starting point, we show the exact solutions for linear JR states and trivial states which have already obtained in \cite{tranjr}.  Then, in Section 3, we investigate the profiles of these two localized states with cubic-quintic nonlinearity. In Section 4, we focus on the detunings of these nonlinear localized states. Finally, in Section 5 we summarize our results and finish with concluding remarks.

\section{GOVERNING EQUATIONS AND LINEAR SOLUTIONS OF JACKIW-REBBI STATES AND TRIVIAL STATES}
\label{linearcase}

As a starting point, in this Section let us briefly introduce the governing equations in interfaced BWAs with cubic-quintic nonlinearity and the \emph{linear} exact solutions of JR states and trivial states which have already been obtained in \cite{tranjr}. This is necessary because some results of linear JR and trivial solutions will be needed later for discussion of nonlinear localized states.

Light propagation in a discrete, periodic binary array of waveguides with cubic-quintic nonlinearity can be described in the continuous wave regime by the following dimensionless coupled-mode equations (CMEs):
\eq{CWCM}{i\frac{da_{n}}{dz}+\kappa[a_{n+1}+ a_{n-1}] - (-1)^{n} \sigma a_{n} +  \gamma(1-b_{s}|a_{n}|^{2}) |a_{n}|^{2}a_{n}=0,}
where $a_{n}$ is the electric field amplitude in the $n$th waveguide, $z$ is the longitudinal spatial coordinate, $2\sigma$ and $\kappa$ are the propagation mismatch and the coupling coefficient between two adjacent waveguides of the array, respectively, $\gamma$ is the nonlinear coefficient of the cubic terms of waveguides, and $b_{s}$ is the saturation parameter governing the power level at which the nonlinearity begins to saturate. For many materials $b_{s}|a_{n}|^{2} \ll 1$ in most practical situations. However, this term may become relevant when the peak intensity approaches 1 GW/$cm^{2}$ in the case of silica \cite{agrawal}. Note that the cubic-quintic nonlinearity is also called competing nonlinearities and $b_{s}$ is always positive for most of optical materials such as semiconductor waveguides, semiconductor-doped  glasses, and organic polymers \cite{kivshar}. It is also worth mentioning that the cubic-quintic nonlinearity is a special case of a more general kind of nonlinearity known as saturable nonlinearity which exists in many nonlinear media (see Eq. (7.4.1) in \cite{kivshar}).

\begin{figure}[htb]
  \centering \includegraphics[width=0.45\textwidth]{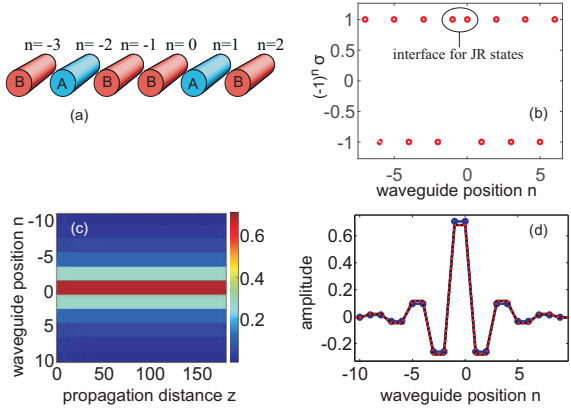}
\caption{\small{(Color online) (a) The scheme of two adjacent BWAs. (b) The value distribution of the array $(-1)^n\sigma$ which can support the discrete JR states. (c) Propagation of a beam in BWAs in the linear regime where Eq. (\ref{solutioncontinuous}) is used as input condition. (d) Curves showing the amplitudes of the beam at input (solid blue with round markers) and output (solid black) which is hidden behind the dotted red curve representing the discrete solution in the form of Eq.  (\ref{solutiondiscrete1}). Parameters are $-\sigma_{1} = \sigma_{2} = 1$; $\kappa = 1$; $\gamma = 0$. Figs. 1(a,c,d) are reproduced from Fig. 1 in \cite{tranjr}.}}
  \label{fig1}
\end{figure}

In order to generate JR states one needs to use two BWAs which are placed adjacent to each other as schematically shown in Fig. \ref{fig1}(a). We want to emphasize that if $n<0$ (for the left-hand side BWA) $\sigma$ takes the constant value $\sigma_{1}$, whereas if $n\geq0$ (the right-hand side BWA) $\sigma$ takes the constant value $\sigma_{2}$.

After setting $\Psi_{1}(n) = (-1)^{n}a_{2n}$ and $\Psi_{2}(n) = i(-1)^{n}a_{2n-1}$, and following the standard approach developed in \cite{zitterbewegung,longhi2} we can introduce the continuous transverse coordinate $\xi \leftrightarrow n$ and the  two-component spinor $\Psi(\xi,z)$ = $(\Psi_{1},\Psi_{2})^{T}$ which satisfies the 1D nonlinear Dirac equation:
\eq{diracequation}{i\de_{z}\Psi = -i\kappa\hat{\sigma}_{x}\de_{\xi}\Psi + \sigma\hat{\sigma}_{z}\Psi - \gamma G + \gamma b_{s} F,} where the cubic nonlinearity is taken into account via the term $G \equiv (|\Psi_{1}|^{2}\Psi_{1},|\Psi_{2}|^{2}\Psi_{2})^{T}$;  the quintic nonlinearity is taken into account via the term $F \equiv (|\Psi_{1}|^{4}\Psi_{1},|\Psi_{2}|^{4}\Psi_{2})^{T}$; $\hat{\sigma}_{x}$ and $\hat{\sigma}_{z}$ are the usual Pauli matrices. In quantum field theory the parameter $\sigma$ in the Dirac equation is often called the mass of the Dirac field (or Dirac mass), and this mass parameter can be both positive and negative. In the case of just Kerr nonlinearity the resulting equation will be simplified as Eq. (7) in \cite{trandirac1} without the quintic term.

In the linear case (i.e., when $\gamma$ = 0), if $\sigma_{1}<0$ and $\sigma_{2}>0$ the exact continuous JR solutions of Eq. (\ref{diracequation}) have already been derived in \cite{tranjr} as follows:
\eq{solutioncontinuous}{ \Psi(\xi) = \sqrt{\frac{|\sigma_{1}\sigma_{2}|}{\kappa(|\sigma_{1}|+|\sigma_{2}|)}}\left(\begin{array}{cc} 1 \\ i \end{array}\right)e^{-|\sigma(\xi)\xi|/\kappa}.} Note that solution (\ref{solutioncontinuous}) is the exact one to  the continuous Eq. (\ref{diracequation}) in the linear case, but it is an approximate solution to the discrete Eqs. (\ref{CWCM}). Obviously, this approximation will become better if the beam width gets larger and larger.

If $|\sigma_{1}|=|\sigma_{2}|=\sigma_{0}$, as shown in \cite{tranjr}, one can easily get exact localized solutions for the discrete Eqs. (\ref{CWCM}) without nonlinearity ($\gamma$ = 0) for the following two cases:

If $-\sigma_{1}=\sigma_{2}=\sigma_{0}>0$, one gets the following discrete JR state \cite{tranjr}:
\eq{solutiondiscrete1}{a_{n} = b_{n}e^{i\delta_{1}z},} where the detuning $\delta_{1} \equiv \kappa-\sqrt{\sigma_{0}^{2}+\kappa^{2}}$, $b_{n}$ is real and independent of the variable $z$, $b_{2n-1} = b_{2n}$. If $n\geq0$ one has the following relationship: $b_{2n}/b_{2n+1} = \alpha \equiv -[\sigma_{0}/\kappa + \sqrt{1+\sigma_{0}^{2}/\kappa^{2}}]$, whereas for $n<0$ one has: $b_{2n+1}/b_{2n} = \alpha$. Note that for generating a discrete JR state, two adjacent waveguides at the interface must have positive values for  $(-1)^{n}\sigma$ $(-1)^{n}\sigma$ (see Fig. \ref{fig1}(b) at the central region for more details). We want to stress that from this discrete solution $b_{n}$ one can easily construct the component $\Psi_{1}(n)$ which has the \emph{same} sign for all $n$ (of course, after dropping the common factor $e^{i\delta_{1}z}$). This is also true for $\Psi_{2}(n)$. Therefore, the derivative $\de_{\xi}\Psi$ in Eq. (\ref{diracequation}) mathematically makes sense in this case, and the discrete solution (\ref{solutiondiscrete1}) can be claimed as the approximate discrete JR solution to the continuous Dirac equation (\ref{diracequation}).

However, if $\sigma_{1}=-\sigma_{2}=\sigma_{0}>0$, one has the following trivial localized state \cite{tranjr}:
\eq{solutiondiscrete2}{a_{n} = b_{n}e^{i\delta_{2}z},} where the detuning $\delta_{2} \equiv \kappa+\sqrt{\sigma_{0}^{2}+\kappa^{2}}$, $b_{n}$ is again real and independent of the variable $z$, $b_{2n-1} = b_{2n}$. If $n\geq0$ one has: $b_{2n}/b_{2n+1} = -\alpha$, whereas for $n<0$ one has: $b_{2n+1}/b_{2n} = -\alpha$. Note that for generating this trivial state, two adjacent waveguides at the interface must have negative values for $(-1)^{n}\sigma$. Here we want to emphasize that from this discrete solution $b_{n}$ one can easily construct the component $\Psi_{1}(n)$ which has \emph{opposite} signs for neighboring values of $n$ (of course, after dropping the common factor $e^{i\delta_{1}z}$), i.e., $\Psi_{1}(n).\Psi_{1}(n+1) \leq 0$ for all $n$. This is also true for $\Psi_{2}(n)$. Therefore, the derivative $\de_{\xi}\Psi$ in Eq. (\ref{diracequation}) mathematically does not make sense in this case, and the discrete solution (\ref{solutiondiscrete2}), unlike the solution (\ref{solutiondiscrete1}), cannot be claimed as the approximate JR solution to the continuous Dirac equation (\ref{diracequation}). That is the reason why we here explicitly call Eqs. (\ref{solutiondiscrete2}) a trivial localized solution to Eqs. (\ref{CWCM}).

In Fig. \ref{fig1}(b) we show the value distribution of the array $(-1)^n\sigma$ for the interfaced BWAs which can support the discrete JR states. In Fig. \ref{fig1}(c) we show the propagation of a beam in the linear regime where the continuous JR solution in the form of Eq. (\ref{solutioncontinuous}) is used as the initial condition for numerically solving the discrete Eqs. (\ref{CWCM}). In Fig. \ref{fig1}(d) we plot the input (the solid blue curve with round markers which is constructed by using the continuous JR solution (\ref{solutioncontinuous})) and output beam amplitudes (the solid black curve) taken from Fig. \ref{fig1}(c). In Fig. \ref{fig1}(d) we also plot the dotted red curve representing the exact discrete JR solution in the form of Eqs. (\ref{solutiondiscrete1}) for the discrete model represented by Eqs. (\ref{CWCM}). The dotted red curve and the output solid black curve in Fig. \ref{fig1}(d) coincide perfectly with each other, therefore the output solid black curve is completely hidden behind the dotted red curve. This fact demonstrates that the continuous JR solution (\ref{solutioncontinuous}) is an excellent approximate solution for the discrete Eqs. (\ref{CWCM}). Similarly, the discrete JR solution (\ref{solutiondiscrete1}) is also an excellent approximate solution for the continuous Dirac Eq. (\ref{diracequation}). In other words, this fact confirms that the discrete solution (\ref{solutiondiscrete1}) can be claimed as a true JR state solution.

To estimate real physical parameters of the calculated DS solitons below we use typical parameters in waveguide arrays made of AlGaAs \cite{morandotti2}, where the coupling coefficient and nonlinear coefficient in physical units are $K=1240m^{-1}$ and $\Gamma=6.5m^{-1}W^{-1}$, respectively. In this case, the power scale will be $P_{0}=K/\Gamma = 190.8 W$, and the length scale in the propagation direction will be $z_{0} = 1/K = 0.8 mm$.

\section{LOCALIZED JACKIW-REBBI STATES AND TRIVIAL STATES WITH CUBIC-QUINTIC NONLINEARITY}
\label{cubicquintic}

Now it is time for us to investigate the nonlinear localized solutions in the presence of cubic-quintic nonlinearity. Specifically, we will solve the nonlinear discrete Eqs. (\ref{CWCM}). Like in the case of Kerr nonlinearity investigated in \cite{tranjr2}, we will find the nonlinear localized solutions in the following form:
\eq{nonlinearsolution}{a_{n} = b_{n}e^{i\delta z},} where the amplitude $b_{n}$ is real and independent of the variable $z$ as in solutions (\ref{solutiondiscrete1}) and (\ref{solutiondiscrete2}). The detuning $\delta$ is the eigenvalue of each nonlinear localized state and must be found further. Obviously, in the linear limit the detuning $\delta$ will get the constant value of either $\delta_{1}$ or $\delta_{2}$ depending on whether one has the discrete JR state or trivial localized state, respectively. After inserting the ansatz (\ref{nonlinearsolution}) into the coupled-mode equations (\ref{CWCM}) we will get the following system of algebraic cubic-quintic equations:

\eq{algebraeq35}{-\delta b_{n} = - \kappa[b_{n+1}+ b_{n-1}] + (-1)^{n} \sigma b_{n} -  \gamma(1-b_{s}|b_{n}|^{2}) |b_{n}|^{2}b_{n}.}

\begin{figure}[htb]
  \centering \includegraphics[width=0.45\textwidth]{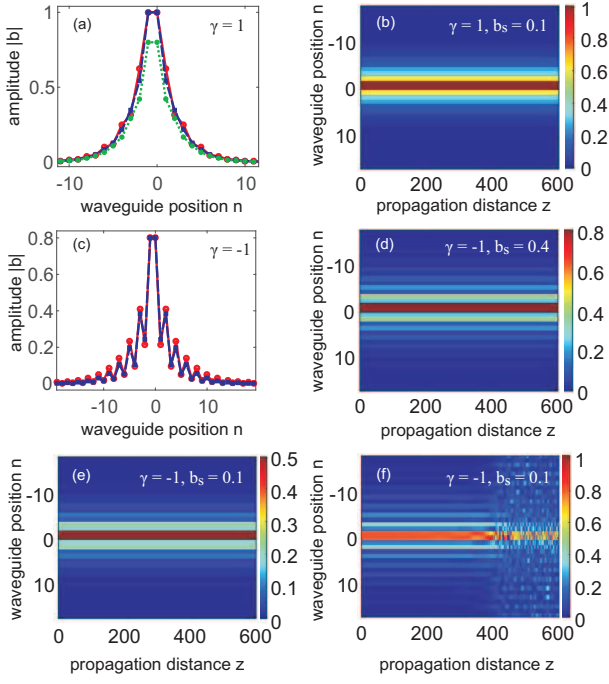}
  \caption{\small{(Color online) Localized discrete JR states in interfaced BWAs with cubic-quintic nonlinearity. (a) Amplitude profile $|b_{n}|$ when $\gamma = 1$: the red curve with round markers is when $b_{s}$ = 0.1 and $b_{0}$ = 1, the dashed blue curve is when $b_{s}$ = 0.4 and $b_{0}$ = 1, whereas the dotted green curve is when $b_{s}$ = 0.4 and $b_{0}$ = 0.8. (b) Propagation of the nonlinear JR state with input condition taken from the red curve with round markers in (a). (c) Amplitude profile $|b_{n}|$ when $\gamma = -1$ and $b_{0}$ = 0.8: the red curve with round markers is when $b_{s}$ = 0.1, whereas the dashed blue curve is when $b_{s}$ = 0.4. (d) Propagation of the nonlinear JR state with input condition taken from the dashed blue curve in (c). (e) Propagation of the nonlinear JR state with $\gamma$ = -1 and $b_{0}$ = 0.5. (f) Propagation of the nonlinear JR state with input condition taken from the red curve with round markers in (c). Parameters: $\sigma_{1}$ = -1; $\sigma_{2}$ = 1; $\kappa$ = 1.}}
  \label{fig2}
\end{figure}

Now we can use certain numerical methods, such as the shooting method to numerically solve Eqs. (\ref{algebraeq35}) for obtaining the nonlinear localized solutions with cubic-quintic nonlinearity and their detuning values. This shooting method \cite{shooting} has been described in details in \cite{tranjr2} to find localized solutions with Kerr nonlinearity. This method can also be successfully used for any other kind of nonlinearity, including the cubic-quintic one as in this work. The essence of this method is that for each fixed value of the peak amplitude $b_{0} = b_{-1}$ one can find the eigenvalue (if it exists) of the detuning $\delta$ such that the tails of the nonlinear localized states will vanish when $n \rightarrow \pm\infty$, i.e., $b_{n} \rightarrow 0$ when $n$ is large enough \cite{tranjr2}. In Fig. \ref{fig2}(a) we plot the amplitude profile $|b_{n}|$ of the localized discrete JR state when $\gamma = 1$. In Fig. \ref{fig2}(a) the red curve with round markers is when $b_{s} = 0.1$, $b_{0}$ = 1 (the eigenvalue of the detuning is found to be $\delta \simeq -0.6682 \delta_{1}$); the dashed blue curve is when $b_{s} = 0.4$, $b_{0}$ = 1 (the eigenvalue of the detuning is found to be $\delta \simeq -0.1360 \delta_{1}$); and the dotted green curve is when $b_{s} = 0.4$, $b_{0}$ = 0.8 (the eigenvalue of the detuning is found to be $\delta \simeq 0.1183 \delta_{1}$). In Fig. \ref{fig2}(b) we show the propagation of the nonlinear localized JR state in the (n,z)-plane by solving Eqs. (\ref{algebraeq35}) with the initial conditions taken from the red curve with round markers in Fig. \ref{fig2}(a), i.e., when $\gamma = 1$, $b_{0}$ = $b_{-1} = 1.0$, and $b_{s} = 0.1$. It is clearly shown in Fig. \ref{fig2}(b) that the profile of this nonlinear JR state is perfectly conserved during propagation for a very long distance. In Fig. \ref{fig2}(c) we plot the amplitude profiles $|b_{n}|$ of the localized discrete JR state in the case of defocusing nonlinearity ($\gamma = -1$) with peak amplitude $b_{0}$ = $b_{-1} = 0.8$ and two values of the saturation parameter $b_{s}$: the red curve with round markers is when $b_{s} = 0.1$ (the eigenvalue of the detuning is found to be $\delta \simeq 2.095 \delta_{1}$), and the dashed blue curve is when $b_{s} = 0.4$ (the eigenvalue of the detuning is found to be $\delta \simeq 1.88 \delta_{1}$). The latter curve is used as initial conditions for solving Eqs. (\ref{algebraeq35}), and the propagation in the (n,z)-plane of this JR state is demonstrated in Fig. \ref{fig2}(d). In Fig. \ref{fig2}(e) we show the propagation of the nonlinear localized discrete JR state in the (n,z)-plane when $\gamma = -1$, $b_{0}$ = $b_{-1} = 0.5$, and $b_{s} = 0.1$ (the eigenvalue of the detuning is found to be $\delta \simeq 1.4428 \delta_{1}$). Meanwhile, in Fig. \ref{fig2}(f) we show the propagation of the nonlinear discrete JR state with all parameters as in Fig. \ref{fig2}(e) except for the input peak amplitude $b_{0}$ = $b_{-1} = 0.8$ in Fig. \ref{fig2}(f) (the eigenvalue of the detuning is found to be $\delta \simeq 2.095 \delta_{1}$) instead of the value $b_{0}$ = $b_{-1} = 0.5$ in Fig. \ref{fig2}(e). In Fig. \ref{fig2} we fix the value of the Dirac masses $\sigma_{1} = -1$, $\sigma_{2} = 1$, and the coupling coefficient $\kappa = 1$.

\begin{figure}[htb]
  \centering \includegraphics[width=0.45\textwidth]{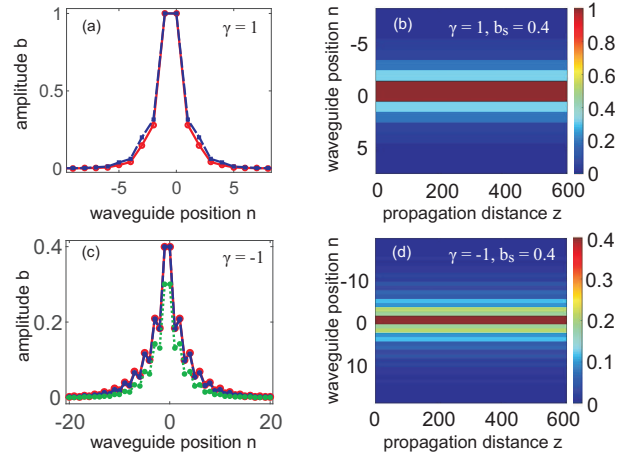}
  \caption{\small{(Color online). Trivial localized states in interfaced BWAs with cubic-quintic nonlinearity. (a) Amplitude profile $b_{n}$ when $\gamma = 1$ and $b_{0}$ = 1: the red curve with round markers is obtained with $b_{s}$ = 0.1, whereas the dashed blue curve is obtained with $b_{s}$ = 0.4. (b) Propagation of the nonlinear trivial state with input condition taken from the blue dahsed curve in (a). (c) Amplitude profiles when $\gamma = -1$: the red curve with round markers is obtained when $b_{s}$ = 0.1 and $b_{0}$ = 0.4, the dashed blue curve is obtained when $b_{s}$ = 0.4 and $b_{0}$ = 0.4, whereas the dotted green curve is obtained when $b_{s}$ = 0.4 and and $b_{0}$ = 0.3. (d) Propagation of the nonlinear trivial state in the (n,z)-plane with input condition taken from the dashed blue curve in (c). Parameters: $\sigma_{1}$ = 1; $\sigma_{2}$ = -1; $\kappa$ = 1.}}
  \label{fig3}
\end{figure}

In Fig. \ref{fig3} we show the trivial localized states in the regime of cubic-quintic nonlinearity. In Fig. \ref{fig3}(a) we plot the amplitude profiles $b_{n}$ of the trivial state having the peak amplitude $b_{0}$ = $b_{-1} = 1.0$ in the regime when $\gamma = 1$ with two values of the saturation parameter $b_{s}$: the red curve with round markers is obtained when $b_{s} = 0.1$ and the eigenvalue of the detuning is found to be $\delta \simeq 1.3172 \delta_{2}$, whereas the dashed blue curve are obtained when $b_{s} = 0.4$ and the eigenvalue of the detuning is found to be $\delta \simeq 1.2071 \delta_{2}$. These trivial states are also very robust and can propagate without any distortion of their shapes for quite a long distance. Indeed, as an example, in Fig. \ref{fig3}(b) we demonstrate the evolution in the (n,z)-plane of the trivial state with input condition taken from the dashed blue curve in Fig. \ref{fig3}(a). Similarly, in Fig. \ref{fig3}(c) we plot the amplitude profiles $b_{n}$ of the trivial states when $\gamma = -1$: the red curve with round markers is obtained when $b_{s} = 0.1$, $b_{0}$ = 0.4 (the eigenvalue of the detuning is found to be $\delta \simeq 0.9527 \delta_{2}$); the dashed blue curve is got when $b_{s} = 0.4$, $b_{0}$ = 0.4 (the eigenvalue of the detuning is found to be $\delta \simeq 0.9548 \delta_{2}$); whereas the dotted green curve is got when $b_{s} = 0.4$, $b_{0}$ = 0.3 (the eigenvalue of the detuning is found to be $\delta \simeq 0.9735 \delta_{2}$). The dashed blue curve in Fig. \ref{fig3}(c)  is used as initial input conditions for investigating the trivial state propagation in the (n,z)-plane shown in Fig. \ref{fig3}(d).

We can see that there are some common features between the profiles of localized states with cubic-quintic nonlinearity shown in Fig. \ref{fig2}, Fig. \ref{fig3} and the ones of localized states with Kerr nonlinearity reported in \cite{tranjr2}. First, if $\gamma$ is positive, then all the profiles of nonlinear localized states monotonically decrease from the centers to the tails, whereas this is not the case if $\gamma$ is negative. Second, if we fix all parameters of localized states except for the sign of $\gamma$, then the profiles of localized states with positive $\gamma$ are more localized in the transverse direction than the ones with negative $\gamma$. Apart from these common points, the quintic term with the saturation parameter $b_{s}$ leads to some new interesting features. First and foremost, $b_{s}$ can help to stabilize localized states and suppress the noise during propagation, especially in the regime of self-defocusing nonlinearity for the cubic term (i.e., when $\gamma$ = -1). Indeed, in Fig. \ref{fig2}(d) and Fig. \ref{fig2}(f) as an example we show the propagation of discrete JR states with all the same parameters except for the only difference that the saturation parameter $b_{s}$ = 0.4 in Fig. \ref{fig2}(d), but $b_{s}$ = 0.1 in Fig. \ref{fig2}(f). The discrete JR state with greater $b_{s}$ shown in Fig. \ref{fig2}(d) is perfectly stable during propagation. Meanwhile, the discrete JR state with smaller $b_{s}$ shown in Fig. \ref{fig2}(f) is only stable at the beginning up to the distance $z \simeq 300$, after that the noise grows up and totally destroys the discrete JR state at the output. So, one can say that media with larger saturation parameter $b_{s}$ is more favorable to support the stable propagation of localized states. This is expected because the saturable nonlinearity is well known for being able to arrest the pulse collapse and helping to form a stable light bullet (a spatiotemporal soliton) in bulk media \cite{kivshar}. Note that the saturation parameter is more crucial in stabilizing localized states with negative $\gamma$. In case of positive $\gamma$ our simulations show that all localized states that we have obtained are quite robust even without saturation parameter $b_{s}$. This is also in line with the results obtained in the case of just Kerr nonlinearity in \cite{tranjr2} (see Fig. 3 therein) which show that media with positive $\gamma$ can support stable localized states for a wide range of peak amplitudes, whereas media with negative $\gamma$ are only able to support localized states with low peak intensity.

Note also that one can make a localized state with negative $\gamma$ more stable by decreasing its initial peak amplitude as shown in Fig. \ref{fig2}(e) where all parameters are exactly the same as in Fig. \ref{fig2}(f) except for a lower initial peak amplitude $b_{0}$ ($b_{0}$ = 0.5 in Fig. \ref{fig2}(e), but $b_{0}$ = 0.8 in Fig. \ref{fig2}(f)). Physically, the mechanism for stabilizing localized states in this case is similar to the one just mentioned above when $b_{s}$ is increased. Indeed, as pointed out in \cite{tranjr2} where localized states with just Kerr nonlinearity have been analyzed and mentioned just above, media with negative $\gamma$ can only support localized states with low peak amplitudes. In other words, in self-defocusing media with Kerr nonlinearity (i.e., when $\gamma < 0$), localized states can be supported only in the case when the nonlinear term is small in absolute value. This is understandable, because in the case of Kerr nonlinearity with $\gamma < 0$, the nonlinear term leads to the \emph{defocusing}, i.e., broadening of the beams. That is why if $\gamma < 0$ one needs to keep the nonlinear term small in absolute value if one wants to create a favorable condition for generating localized states. In the case of cubic-quintic nonlinearity with $\gamma < 0$, because $b_{s}$ is always positive and $1 - b_{s}|b_{n}|^{2} > 0$, one can decrease the influence of the nonlinear term in Eqs. (\ref{algebraeq35}) by two ways: either by decreasing the peak amplitude $b_{0}$, or by increasing the saturation parameter $b_{s}$. So, both these ways will be able to support stable nonlinear localized states in the case of cubic-quintic nonlinearity as shown in Figs. \ref{fig2}(d,e).

It is worth emphasizing that profiles of localized states with Kerr nonlinearity reported in \cite{tranjr2} have a very interesting feature: their peak amplitude and their transverse dimension increase (or decrease) at the same time. Our results in this work show that this distinguishing feature of localized states in BWAs is universal not only for localized states with Kerr nonlinearity, but also for localized states with cubic-quintic nonlinearity. Indeed, in Fig. \ref{fig2}(a) the dashed blue curve and the dotted green curve of discrete JR states are obtained when all parameters are exactly the same except for the peak amplitude. We see that the dotted green curve with lower peak amplitude is totally enveloped by the dashed blue curve with higher peak amplitude. The same situations happens for the dashed blue curve and the dotted green curve of trivial states in Fig. \ref{fig3}(c). Meanwhile, for all other well-known localized structures (including optical solitons emerging from the famous nonlinear Schr\"{o}dinger equation, Bragg solitons, discrete solitons in a conventional waveguide array (see \cite{kivshar} for more details), and even discrete Dirac solitons in a BWA found in \cite{trandirac1}) these two characteristic parameters always vary in an opposite manner: the peak increase will lead to the spatial narrowing of the beam (or the temporal shortening of the pulse) and vice versa.

\section{DETUNINGS OF JACKIW-REBBI STATES AND TRIVIAL STATES WITH CUBIC-QUINTIC NONLINEARITY}
\label{detunings}

\begin{figure}[htb]
  \centering \includegraphics[width=0.45\textwidth]{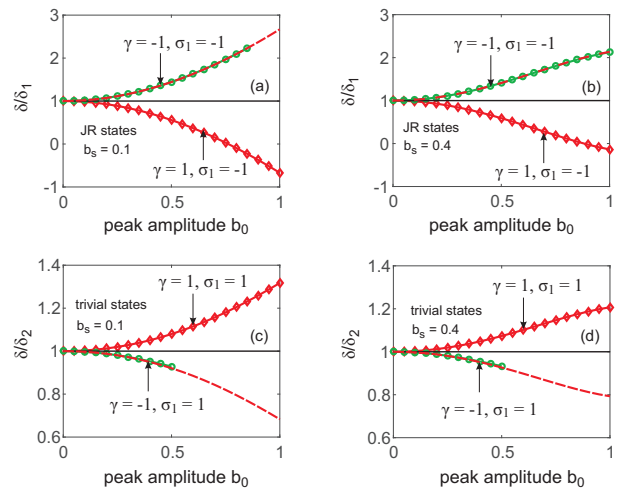}
  \caption{\small{(Color online) Relative detuning of nonlinear localized states as functions of the peak amplitude $b_{0}$ in the case of cubic-quintic nonlinearity. (a,b) The curves are obtained for discrete JR states with the saturation parameter $b_{s}$ = 0.1 in (a) and $b_{s}$ = 0.4 in (b). (c,d) Exactly the same as in (a,b), respectively, but for trivial states. The dashed red curves in Figs. \ref{fig4} are the mirror images of the red curves with diamond markers with respect to the black horizontal axis. The coupling coefficient $\kappa = 1$, as usual. All other parameters are indicated for each curve therein.}}
  \label{fig4}
\end{figure}

In this Section we investigate in details the dependence of the localized states detunings on their peak amplitudes and other parameters. In Fig. \ref{fig4}(a) we plot the relative detuning $\delta/\delta_{1}$ of nonlinear discrete JR states as functions of the peak amplitude $b_{0}$ in the case of cubic-quintic nonlinearity with the saturation parameter $b_{s}$ = 0.1. The red curve with diamond markers in Fig. \ref{fig4}(a) is obtained when $\gamma$ = 1, $\sigma_{1}$ = -1, and $\sigma_{2}$ = 1; whereas the green curve with round markers is obtained when $\gamma$ = -1, $\sigma_{1}$ = -1, and $\sigma_{2}$ = 1. These two curves are almost symmetrical with respect to the black horizontal axis which represents the detuning of the linear discrete JR states. Indeed, in Fig. \ref{fig4}(a) the dashed red curve (which is the mirror image of the red curve with diamond markers with respect to the black horizontal axis) coincides very well with the green curve with round markers. There is only one significant difference between these two curves: the one with positive $\gamma$ can be drawn further for higher peak amplitude $b_{0}$, whereas the other one with negative $\gamma$ stops at the maximum value $b_{0} \simeq 0.85$, and we cannot find localized solutions for JR states with higher value $b_{0}$ in this case. The symmetry of the two curves representing detunings of localized states in interfaced BWAs with just Kerr nonlinearity while switching the sign of $\gamma$ and keeping all other parameters fixed has been explained in details in Section 4 in \cite{tranjr2}. This explanation is also valid for localized states in interfaced BWAs of all other types of nonlinearity, including the cubic-quintic term as in this work. Indeed, now we follow the very reasoning explained in details in Section 4 in \cite{tranjr2} with the only exception that we have the cubic-quintic nonlinearity in this work instead of Kerr nonlinearity in \cite{tranjr2}. Suppose that $\delta_{l}$ is the detuning parameter of {\em linear} localized states, i.e., $\delta_{l}$ = $\delta_{1}$ for linear JR states, and $\delta_{l}$ = $\delta_{2}$ for trivial states. Suppose also that $\delta$ is the detuning of the nonlinear localized states at the peak amplitude $b_{0}$ in the case of $\gamma$. Now we switch the sign of $\gamma$ so that $\gamma \rightarrow -\gamma$. By doing that the detuning of the nonlinear localized states will be transformed $\delta \rightarrow \delta^{'}$, and the amplitude will be changed $b_{n} \rightarrow c_{n}$. In order to have the situation when the curve representing $\delta/\delta_{l}$ is almost the mirror image of the curve representing $\delta^{'}/\delta_{l}$ with respect to the black horizontal line in Fig. \ref{fig4}, one must have the following relationship: $\delta/\delta_{l} \simeq 2 - \delta^{'}/\delta_{l}$. As a result, one has: $\delta^{'} \simeq 2\delta_{l} - \delta$. So, for the case of $-\gamma$, Eqs. (\ref{algebraeq35}) will now have the following form:

\eq{algebraeq2}{
\begin{split}
-(2\delta_{l} - \delta)c_{n} \simeq - &\kappa[c_{n+1}+ c_{n-1}] + (-1)^{n} \sigma c_{n} +  \\
                                      &\gamma(1-b_{s}|c_{n}|^{2}) |c_{n}|^{2}c_{n}.
\end{split}
}
Now we add Eqs. (\ref{algebraeq35}) to Eqs. (\ref{algebraeq2}) and get the following equations:

\eq{algebraeq3}{
\begin{split}
&\delta (c_{n} - b_{n}) -2\delta_{l}c_{n} \simeq - \kappa[(c_{n+1}+b_{n+1}) + (c_{n-1} + b_{n-1})] + \\
&(-1)^{n} \sigma (c_{n} + b_{n}) +  \gamma (|c_{n}|^{2}c_{n} - |b_{n}|^{2}b_{n}) + \\
& \gamma b_{s} (|b_{n}|^{4}b_{n} - |c_{n}|^{4}c_{n}).
\end{split}
} As in \cite{tranjr2}, now it is easy to see that if we fix the peak amplitudes so that $b_{0} = c_{0}$ and suppose that the condition $c_{n} \simeq b_{n}$ is held true (which is practically satisfied if the peak amplitudes are small, i.e., when we operate in the regime close to the linear case), then all the nonlinear terms in Eqs. (\ref{algebraeq3}) will practically vanish. As a result, from Eqs. (\ref{algebraeq3}) we will have:

\eq{algebraeq4}{-\delta_{l}c_{n} \simeq - \kappa[c_{n+1}+ c_{n-1}] + (-1)^{n} \sigma c_{n}.}
The latter equations are again automatically satisfied in the regime close to the linear case, because parameter $\delta_{l}$ in Eqs. (\ref{algebraeq4}) is the detuning of the {\em linear} localized states as set above (see also Eqs. (\ref{algebraeq35}) in the linear regime when $\gamma$ = 0).

So, in interfaced BWAs with cubic-quintic nonlinearity we have proved that when changing the sign of the nonlinear coefficient $\gamma$ and fixing all other parameters one will also obtain two curves representing relative detunings of nonlinear localized states which are almost symmetrical with respect to the horizontal axis representing the detuning of the corresponding \emph{linear} localized states. Note that, as mentioned in \cite{tranjr2}, this symmetry is satisfied very well when the peak amplitude of nonlinear localized states is small, but the violation of this symmetry will be more and more significant if we increase the peak amplitude of nonlinear localized states.

In Fig. \ref{fig4}(b) we plot the same quantities as in Fig. \ref{fig4}(a) with the only difference that now $b_{0}$ = 0.4. One can see that the larger value of the saturation parameter $b_{s}$ in Fig. \ref{fig4}(b) helps to develop the green curve with round markers further for higher values of the peak amplitude $b_{0}$. This fact is possible because if $\gamma < 0$, then the quintic term with $b_{s}$ in Eqs. (\ref{algebraeq35}) acts as the self-focusing nonlinearity which helps to balance the effect of the self-defocusing nonlinearity of the cubic term for higher peak amplitudes. Note that, as pointed out in \cite{tranjr2} and mentioned above, the self-focusing nonlinearity is always more favorable to support high-intensity JR states than the self-defocusing nonlinearity.

In Figs. \ref{fig4}(c,d) we plot the relative detuning $\delta/\delta_{2}$ of nonlinear trivial states as functions of the peak amplitude $b_{0}$. All parameters and curves in Figs. \ref{fig4}(c,d) are exactly similar to the ones in Figs. \ref{fig4}(a,b), respectively, with the only difference that Figs. \ref{fig4}(a,b) represent the case of nonlinear discrete JR states, whereas Figs. \ref{fig4}(c,d) represent the case of nonlinear trivial states (as a result, $\sigma_{1}$ must be negative in Figs. \ref{fig4}(a,b), but positive in Figs. \ref{fig4}(c,d)). One can see from Figs. \ref{fig4}(c,d) that it is also possible to obtain trivial states with high peak amplitudes in the regime of cubic-quintic nonlinearity for the case of positive $\gamma$. However, with negative $\gamma$ one can only generate trivial states with rather low peak amplitudes $b_{0} \leq 0.5$ in Figs. \ref{fig4}(c,d). Note that one can get discrete JR states with higher peak amplitudes $b_{0}$ for negative $\gamma$ in Figs. \ref{fig4}(a,b). This situation is quite similar to the case of Kerr nonlinearity reported in \cite{tranjr2}. Note also that the larger value of the saturation parameter $b_{s}$ in Fig. \ref{fig4}(d) does not help to generate trivial states with higher peak amplitudes (two green curves with round markers stop at $b_{0}$ = 0.5 in both Fig. \ref{fig4}(c) with $b_{s}$ = 0.1, and Fig. \ref{fig4}(d) with $b_{s}$ = 0.4). We suppose that the reason is because trivial states with negative $\gamma$ are much less localized in the transverse direction than nonlinear JR states both with the Kerr and cubic-quintic nonlinearity. Therefore, it is always more difficult to obtain the localized nonlinear trivial states with high peak amplitudes when $\gamma$ = -1.

\section{CONCLUSION}
\label{CONCLUSION}

In this work we have systematically investigated two types of localized states - the optical analogues of quantum relativistic Jackiw-Rebbi states and trivial localized states - in interfaced BWAs in the regime of cubic-quintic nonlinearity. We have shown that large values of the saturation nonlinearity parameter can help to generate and stabilize localized states of both types with high peak amplitudes, especially in the case of negative coefficient for the cubic nonlinearity term. Like the profiles of localized states with just Kerr nonlinearity, those of localized states with cubic-quintic nonlinearity in interfaced BWAs possess a distinguishing feature that all other well-known localized nonlinear structures, including solitons, do not have. Namely, the peak amplitude and the transverse dimension of these localized states can increase (or decrease) simultaneously. We have also found out that localized states with the positive sign of the coefficient for the cubic term can exist in a wide range of their peak amplitudes, whereas those with the negative sign can only exist in the low-peak amplitude regime. We have also demonstrated that by changing the sign of the coefficient for the cubic nonlinearity term while keeping all other parameters fixed one can obtain two curves for the relative detunings of nonlinear localized states which are almost symmetrical with respect to the axis representing the detuning of the corresponding linear localized states. This general rule is applicable for all types of localized states with various kinds of nonlinearity in binary waveguide arrays.

\section{ACKNOWLEDGMENTS}
\label{acknowledgments}
This research is funded by Vietnam National Foundation for Science and Technology Development (NAFOSTED) under grant number 103.03-2019.03.

\end{document}